\documentclass[aps,prb,12pt]{revtex4}
\usepackage{graphicx}
\usepackage{amsmath}
\usepackage{amsfonts}
\usepackage{amssymb}
\begin{document}
\title{Phases as Hidden Variables of Quantum Mechanics}
\author{S.V.Gantsevich}
\affiliation{Ioffe Institute of Russian Academy of Sciences,
Saint-Petersburg 194021, e-mail: sergei.elur@mail.ioffe.ru}
\begin{abstract}
\baselineskip=2.5ex {\it Quantum mechanical wave functions have
phases. These phases either initial or acquired during time
evolution usually do not enter the final expressions for
observable physical quantities. Nevertheless in many cases the
observable physical quantities implicitly depend on the phases.
Hence we may regard the phases as a sort of hidden variables of
Quantum Mechanics. Neglecting the phase role makes inexplicable
the peculiar quantum effects such as particle interference,
Einstein-Podolsky-Rosen correlation and many others. To the
contrary the adequate inclusion of phases into consideration
reduces QM puzzles and mysteries to simple and obvious
triviality}.
\end{abstract}
\maketitle \baselineskip=2.5ex Pacs:  03.65.-w, 03.65.Ud\\
\par
\section{Introduction}
Various proofs exist that hidden variables are incompatible with
the principles and rules of Quantum Mechanics (see, e.g.
[\cite{Nmn,Bel,Blf}]). However there are quantities that may be
regarded as a sort of hidden variables of Quantum Mechanics. These
quantities are the phases of wave functions. As a rule these
phases do not enter the final QM expressions for the observed
physical quantities though frequently they implicitly determine
their values or even the very existence of QM effects. Neglecting
the phase role makes inexplicable the peculiar quantum effects
such as particle interference, Einstein-Podolsky-Rosen correlation
and many others. To the contrary the adequate inclusion of phases
into consideration reduces QM puzzles and mysteries to simple and
obvious triviality.
\section{Bra, Ket and Wavicle}
\par
Dirac in his notorious book [\cite {D}] introduced the bra and ket
functions (or vectors in Hilbert space) and the corresponding
expression (bracket) for the observable physical quantity. Let us
consider an observable physical quantity $c$ for the simple case
of a quantum particle in some space volume $V$. In the usual
notations and in Dirac's notation we have:
\begin{equation}\label{1}
\bar{c}=\langle \psi|C|\psi\rangle\equiv\int_V d{\bf r}d{\bf
r'}\psi^\dag({\bf r},t)C({\bf r}|{\bf r'})\psi({\bf r'},t)
\end{equation}
Three quantities stand in this expression: the bra-wave function
(or bra-vector) $\langle \psi|\equiv\psi^\dag({\bf r},t)$ then the
operator of physical observable $\hat{c}$ with the kernel $C({\bf
r}|{\bf r'})$ and the ket-wave function (ket-vector)
$|\psi\rangle\equiv \psi({\bf r'},t)$.
\par
The bra and ket functions are under permanent action of
Hamiltonian $H$ and satisfy the equations of motion
(Schr\"{o}dinger equations). We have for the ket-function
\begin{equation}\label{2}
(\partial_t+iH)\psi(x,t)=\psi(x,0)\delta(t)
\end{equation}
(for brevity we take $\hbar=1$ and do not write the vector signs
for space coordinates ${\bf r}\equiv x$). The solution of the
equation (\ref{2}) in the resolvent form or as the multiplicative
integral is given by
\begin{equation}\label{3}
\psi(x,t)=\frac{1}{\partial_t+iH}\psi(x,0)\delta(t)=\Pi_0^t(1-iHdt)\psi(x,0)
\end{equation}
For the bra-function we have analogously
\begin{equation}\label{4}
\psi^\dag(x,t)=\frac{1}{\partial_t-iH}\psi^\dag(x,0)\delta(t))=\Pi_0^t(1+iHdt)\psi^\dag(x,0)
\end{equation}
The multiplicative integrals at right in (\ref{3}) and (\ref{4})
show how the Hamiltonian acts at the wave function changing it in
every moment of time.
\par
For time-independent $H$ the solutions simplify:
\begin{equation}\label{5}
\psi(x,t)=\exp(-iHt)\psi(x,0), \quad
\psi^\dag(x,t)=\exp(iHt)\psi^\dag(x,0)
\end{equation}
\par
For the eigenfunctions of Hamiltonian $H\psi_p=\epsilon_p\psi_p$
which we take as the set the normalized and orthogonal functions
$$\langle p|k\rangle=0,\quad \langle k|p\rangle=0
 \quad \langle p|p\rangle=1, \quad \langle k|k\rangle=1$$
the time dependence reduces to the phase change
$\psi\equiv\psi_p$:
\begin{equation}\label{6}
\psi^\dag_p(x,t)=e^{+i\epsilon_pt-i\varphi_p}\psi^\dag_p(x) ,\quad
\psi_p(x,t)=e^{-i\epsilon_pt+i\varphi_p}\psi_p(x)
\end{equation}
Here $\varphi_p$ is the initial phase.
\par
Now let us decompose the initial wave function
$\psi(x,0)\equiv\psi_0$ into a series of Hamiltonian
eigenfunctions. Using (\ref{6}) we get for the bra-function
$\psi^\dag(x,t)\equiv \psi^\dag(t)$ and ket-function
$\psi(x,t)\equiv \psi(t)$ the expressions:
\begin{equation}\label{7}
\psi(t)=\sum_pa_pe^{-i\epsilon_pt+i\varphi_p}\psi_p, \quad
\psi^\dag(t)=\sum_pa_p^\dag
e^{+i\epsilon_pt-i\varphi_p}\psi_p^\dag
\end{equation}
The equations of motion for bra and ket are linear so the sum of
solutions is also a solutions. The expressions for physical
observable quantities are bilinear on bra and ket. This
peculiarity of quantum mechanics leads to the existence of two
different contributions to the value of a physical quantity.
Indeed, substituting (\ref{7}) into (\ref{1}) with $C\equiv U(x)$
we obtain two parts of the observable $U$. The first part contains
only diagonal matrix elements $\langle p|U|p\rangle$ of the
operator of observable $U$ and does not depend on time and initial
phases:
\begin{equation}\label{8}
\overline{U}=\sum_p\langle p|U|p\rangle |a_p|^2=\sum_p\langle
p|U|p\rangle F_p
\end{equation}
Here $F_p$ is the occupation number (or distribution function) for
the quantum state $p$. The sum of $F_p$ over $p$ is equal to the
total number $N$ of quantum particles: $N=\sum_pF_p$. For one
particle $F_p$ describes the probability for a particle to be in
the state $p$.
\par
Another part of $U$ contains the non-diagonal matrix elements
$\langle p|U|k\rangle$ and depends on time and initial phases. We
denote it as $\Delta U(t)$:
\begin{equation}\label{9}
\Delta U(t)=\sum_{p\neq k}\langle p|U|k\rangle a^\dag_pa_k
e^{i(\epsilon_p-\epsilon_k)t-i(\varphi_p-\varphi_k)}
\end{equation}
Note that this contribution vanishes after averaging over time or
initial phases.
\par
The total observable $U(t)$ is given by the sum of (\ref{8}) and
(\ref{9}):
\begin{equation}\label{10}
U(t)=\overline{U}+\Delta U(t), \quad \overline{\Delta U(t)}=0
\end{equation}
We see that the physical quantity $U(t)$ is the sum of the
constant background $\overline{U}$ and the alternating
fluctuations $\Delta U(t)$ over this background with zero mean
value. Note that phases explicitly enter only into the fluctuation
part of observable data (\ref{9}). Implicitly the phases are
contained also in the background part (\ref{8}) since the
probability $F_p$ there is formed by the bra and ket with the same
phases. One may say that $F_p$ is the mean number of bra and ket
with the same phases.
\par
Since the operator $U$ can be arbitrary the state of a quantum
object is determined by the bra and ket taken in one moment of
time. It is reasonable to put just such bra+ket pairs (of
functions or objects behind them) into correspondence to the
potentially observable quantum particles. As it is known they are
not waves and not corpuscles revealing nevertheless the wave and
corpuscle properties. Following Eddington let us call such objects
by {\it wavicles}. The bra and ket have phases and the wavicles
also have phases which are equal to the bra and ket phase
differences. The expressions (\ref{8} - \ref{10}) show that there
are zero phase wavicles formed by bra and ket with the same phase.
These wavicles contribute to the constant background of physical
quantities looking like classical particles and revealing
themselves as corpuscles. The wavicles with phases either initial
or acquired during time evolution until the observation moment
reveal themselves as quantum fluctuations. Their contributions
into observable quantities depend on phases and vanish after phase
averaging. Just these wavicles with phases are responsible for the
wave properties of quantum particles.
\section{Diagrams}
\par
Now let us depict all written above as quantum diagrams. The
diagrams are much more transparent than letter formulae. We use
the diagrams where each diagram picture has one to one
correspondence to the letter formulae and one can easily write the
analytical expression for a diagram picture.
\par
The quantum diagram pictures for the wave functions (bra and ket),
the observable quantities, the occupation numbers and the action
of a perturbation potential are presented in the Figure 1.
\begin{figure}[htb]
\begin{center}
\includegraphics[width=4in]{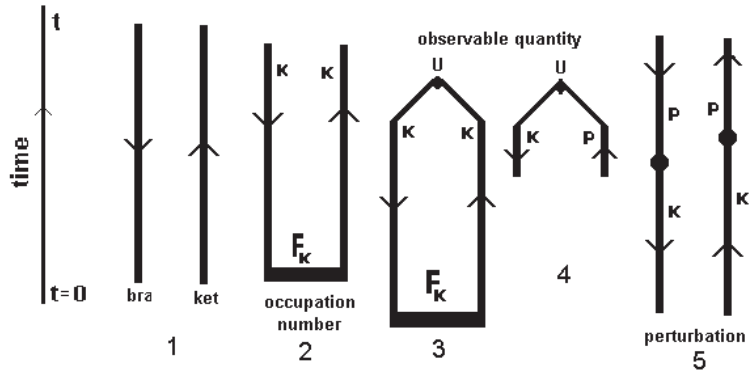}\\
\end{center} \caption{\label{phase1} Basic diagram symbols}
\end{figure}\\
Because of simplicity of these diagrams and their direct
connection with analytical formulae here no rigorous derivation is
necessary. (For more details see [\cite{prb,fnl10,fnl13,gan17}]).
\par
The diagram (F1.1) at the left represents the bra and ket by two
lines with time arrows which correspond $\psi^\dag(t)$ and
$\psi(t)$. The time goes from bottom $t=0$ to the top time $t$. At
$t=0$ we have $\psi^\dag(0)$ for the bra and $\psi(0)$ for the
ket. The line intervals between these points show time evolution
and correspond the resolvents $1/(\partial_t-iH)$ for the bra and
$1/(\partial_t+iH)$ for the ket. Going from top to bottom of the
lines (i.e. against time) and writing consecutively all symbols we
get exactly the solutions of the equations of motion (\ref{3}) and
(\ref{4}). Note that the diagrams represent the solutions of the
equations.
\par
The resolvents can be understood as the infinite series of events
when the Hamiltonian $H$ acts during the time interval $(0t)$ by
zero, one, two, three and more times. For the ket we have:
\begin{equation}\label{11}
\frac{1}{\partial_t+iH}=\frac{1}{\partial_t}+\frac{1}{\partial_t}(-iH)\frac{1}{\partial_t}
+\frac{1}{\partial_t}(-iH)\frac{1}{\partial_t}(-iH)\frac{1}{\partial_t}+...
\end{equation}
and the analogous expression for the bra with the substitution
$(-iH)\rightarrow (+iH)$. The symbols $1/\partial_t$ correspond to
time integrations over intervals where nothing occur while symbols
$H$ describe the momentary Hamiltonian action under which the bra
or ket states abruptly change.
\par
The multiplicative integrals in (\ref{3}) and (\ref{4}) give
another interpretation picture. At each time moment the evolution
looks like some Poisson-type process with the alternative for the
Hamiltonian to act or not to act.
\par
The right diagram (F1.5) describes the action of a perturbation
potential $V$. The potential action can be represented as a series
of events similar to the expression (\ref{11}). We have for the
ket:
\begin{equation}\label{12}
\frac{1}{\partial_t+iH+iV}=\frac{1}{\partial_t+iH}+\frac{1}{\partial_t+iH}(-iV)\frac{1}{\partial_t+iH}
+...
\end{equation}
Here events correspond only to the perturbation action while the
Hamiltonian action is taken into account as the part of normal
evolution. The points on the lines describes the perturbation
actions.  Note that after the perturbation point the bra remains
bra and the ket remains ket. The $k\rightarrow p$ transition under
the action of $V$ gives the usual expression of the perturbation
theory. For the ket we have
\begin{equation}\label{13}
\langle
p|\frac{1}{\partial_t+iH}(-iV)\frac{1}{\partial_t+iH}|k\rangle
=\frac{1}{\partial_t+i\epsilon_p}(-i\langle
p|V|k\rangle)\frac{1}{\partial_t+i\epsilon_k}
\end{equation}
and the similar expression for the bra. Note that in written
formulae the future is at left and the past is at right, so the
ordinary left to right writing order of symbols is against the
diagram time direction.
\par
The diagram (F1.2) shows the time evolution of the wavicle for the
quantum state $k$ occupied with the probability $F_k$. The wavicle
is composed by the pair of bra+ket lines with the same quantum
indices and initial phases. Again going from top to bottom of the
diagram we get:
\begin{equation}\label{14}
F_k(t)=\frac{1}{\partial_t-i\epsilon_k+i\epsilon_k}F_k(0)\delta(t)=F_k(0)\frac{1}{\partial_t}\delta(t)=
F_k(0)\Theta(t)
\end{equation}
Here  $\Theta(t)=(1/\partial_t)\delta(t)$ is the step-function of
Heaviside, $\Theta(t)=0$ for $t<0$ and $\Theta(t)=1$ for $t\geq
0$. The bra and ket of the wavicle have phases which are
time-dependent. But the wavicle phase remains zero because of the
phase equality
$e^{i\epsilon_kt-i\varphi_k}e^{-i\epsilon_kt+i\varphi_k}=1$.
\par
The initial state occupation number $F_k(0)$ is represented by the
horizontal bar at $t=0$. Such symbols and the time ordering of
events (points) on the diagram lines are the main differences
between our diagrams and the widely used Feynman diagrams (see,
e.g. [\cite{F,BD,R}]).
\par
The diagram (F1.3) shows the mean contribution to the physical
quantity $\overline{U}$ from the wavicles of the occupied state
$k$. The value is given by the matrix element $\langle
k|U|k\rangle$ which corresponds to the top point where the bra and
ket lines enter. This point describes the "classical device" of
Bohr needed for a measurement. Measurements really are complicated
physical processes with many stages. All concrete details of such
processes and the necessary averaging of final data are implicitly
included in the values of corresponding matrix elements. Bohr's
"classicality" of a device simply means that two undetectable
elements of Quantum World (i.e. bra and ket) unite in the device
to become detectable in our Classical World.
\par
The contribution to the fluctuation part $\Delta U$ of the
observable $U$ is shown in the diagram (F1.4). This part is given
by the bra and ket with different quantum indices and initial
phases as it is shown in (\ref{9}). Reading the diagram we get
($t\geq t'$):
\begin{equation}\label{15}
\langle
k|U|p\rangle\frac{1}{\partial_t-i\epsilon_k+i\epsilon_p}a^\dag_ka_p
e^{-i(\varphi_k-\varphi_p)}\delta(t-t')=\langle k|U|p\rangle
a^\dag_ka_p
e^{i(\epsilon_k-\epsilon_p)(t-t')-i(\varphi_k-\varphi_p)}
\end{equation}
Let us note that the measurement points in the diagrams (F1.3) and
(F1.4) differ essentially from the perturbation points in the
diagram (F1.5). Perturbations points in bra and ket lines do not
change their properties, i.e. the bra remains bra and ket remains
ket. To the contrary the measurement points where the bra and ket
line unites signify the end of normal (unitary) evolution
prescribed by the motion equations.
\par
The bra and ket lines with different indices can be obtained by a
perturbation action or by the exchange of bra or ket between the
wavicles of occupied states.
\section{Quantum Exchange}
\par
If we take at time $t$ two wavicles of the occupied states $k$ and
$p$ we will get two bra and two ket with two random phases
$\phi_k(t)=\epsilon_kt-\varphi_k$ and
$\phi_p(t)=\epsilon_pt-\varphi_p$. (The phases $\phi(t)$ enter in
phase multipliers as $e^{i\phi(t)}$ for the bra and
$e^{-i\phi(t)}$ for the ket.) From these four {\it independent}
objects of Quantum World we can form four wavicles potentially
observable in our Classical World. Two of them have zero phase
$\beta=0$ while two other have equal and sign-opposite phases
$\alpha(t)$ and $-\alpha(t)$:
\begin{equation}\label{16}
\beta\equiv\phi_p(t)-\phi_p(t)=\phi_k(t)-\phi_k(t)=0, \quad
\alpha(t)\equiv\phi_p(t)-\phi_k(t)=-[\phi_k(t) -\phi_p(t)]
\end{equation}
The phase $\alpha$ is always random even if
$\epsilon_k=\epsilon_p$ and all contributions of such wavicles
into observable quantities are random and vanish after phase
averaging.
\par
Now let us take two independent detectors and find the observable
quantities $A$ and $B$ related to the wavicles of $\beta$ and
$\alpha$ types. Since for the $\alpha$-type wavicles
$\overline{A}=0$ and $\overline{B}=0$ we consider the correlated
values $\overline{AB}$ or $\overline{BA}$. These values taken in
one time moment are depicted in the Figure 2.
\begin{figure}[htb]
\begin{center}
\includegraphics[width=4in]{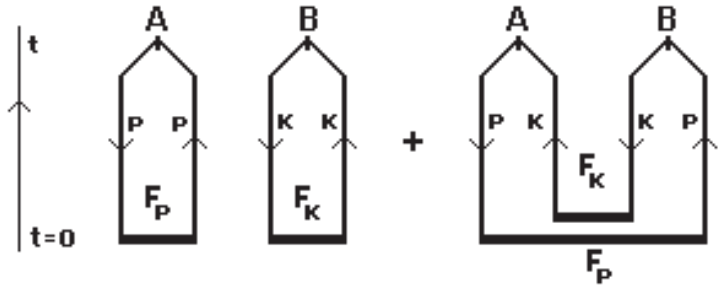}\\
\end{center} \caption{\label{faza2} Exchange correlation}
\end{figure}\\
There are also analogous two diagrams with substitutions
$A\rightarrow B$ and $B\rightarrow A$ which we do not depict for
brevity. The diagrams correspond to the jointly averaged product
of the measurement values in both detectors.
\par
The left diagram shows the contribution of two wavicles with zero
phases. We see that no links between them exist and their
contribution is simply the product of two independent values. One
wavicle hits one detector while other wavicle hits another
detector or vice versa. In this case either separate or joint
averaging of the detector data give the same result. Reading this
diagrams we have (together with analogous one) the uncorrelated
detector contributions:
\begin{equation}\label{17}
(\overline{AB})_{uncor}=[\langle p|A|p\rangle \langle k|B|k\rangle
+ \langle k|A|k\rangle \langle p|B|p\rangle ]F_pF_k
\end{equation}
The right diagram shows the contribution of two wavicles with the
same but sign opposite phases. The detector data of such wavicles
contain the multipliers $\exp(i\alpha)$ and $\exp(-i\alpha)$ and
therefore vanish after separate phase averaging in each detector.
However, after the joint averaging the multipliers cancel each
other and the joint result becomes phase independent. The
corresponding contribution is given by
\begin{equation}\label{18}
(\overline{AB})_{cor}=\pm[\langle p|A|k\rangle\langle
k|B|p\rangle+\langle k|A|p\rangle\langle p|B|k\rangle]F_p F_k
\end{equation}
The total average $\overline{AB}$ is the sum of the uncorrelated
and correlated parts:
\begin{equation}\label{19}
\overline{AB}=[\langle p|A|p\rangle \langle k|B|k\rangle \pm
\langle p|A|k\rangle \langle k|B|p\rangle ]F_p
F_k+(A\rightleftarrows B)
\end{equation}
The sign of correlation contribution is negative for fermions and
positive for bosons.
\par
We see from (\ref{19}) and the diagrams in the Figure 2 how the
exchange of the bra or ket in the zero-phase wavicles produces two
wavicles with the same and sign-opposite phases. These wavicles
become phase correlated and any physical observable quantities
from these wavicles also become correlated. They are just the
so-called entangled quantum particles which are so popular in the
literature. The graphic transformation of the left diagram in the
Figure 2 into the right one inevitably leads to the line
intersection or Schr\"{o}dinger Verschr\"{a}nkung (i.e.
entanglement). The intersection results in the sign difference
between bosons and fermions. (In the Figure 2 the correlation
diagram is depicted without line intersection to make more
transparent the origin of phase correlation).
\par
Let us emphasize that phase correlated (entangled) wavicles have
no magic properties and any action on one of them has no influence
on another. The phase correlation is the real "common cause at the
past" for EPR-correlation, Hunbery Brown-Twiss effect and many
other experimentally observable correlation phenomena.
\par
Note that such phase correlated wavicles are always present in any
many-particle quantum systems. For example in many electron
systems they are responsible for the appearance of exchange
Coulomb energy. Two coherent charge fluctuations there have enough
time to interact by the quick Coulomb potential and give the
corresponding contribution to the energy. Ignoring the phase role
in this effect leads to the usual explanation in QM manuals by
obscure words: "It is a quantum effect with no classical analogy".
\par
In the same way the EPR-correlation on macroscopic distances gave
rise to the multitude of similar "explanations" like nonlocality,
retrocausality, multitude of worlds or other non-physical
fantasies. No such fantasies are needed for the simple physical
picture of quantum exchange correlation. (For the details see
[\cite{gan17,gg17,gg18}]).
\par
The quantum exchange between two wavicles in the same state leads
to the difference in the behavior of bosons and fermions. If we
put $p=k$ in the formula (\ref{19}) we get $AB\equiv 0$ for
fermions since the exchange contribution cancels the product of
independent $A$ and $B$ contributions. In this way the Pauli
prohibition for two fermions to be in the same states realizes.
\par
For the bosons the exchange contribution doubles the uncorrelated
contribution $AB\rightarrow 2AB$. For $A=B=1$ we get the
well-known formula for the occupancy number fluctuations of
independent bosons and fermions:
\begin{equation}\label{20}
\overline{\delta F_p\delta
F_k}\equiv\overline{F_pF_k}-F_pF_k=F_p(1\pm F_p)\delta_{pk}
\end{equation}
Here $F_p\delta_{pk}$ is the Poisson autocorrelation term.
\section{Phase and Probability}
We see above that the phase-independent wavicles (or $\beta$-type
wavicles) form the permanent background of physical quantities.
The phase-dependent wavicles (or $\alpha$-type wavicles) create
fluctuations over this background with mean zero values. Now let
us consider the probabilities of these physical processes. It is
convenient to take as a simple example the mean value of
space-position operator $U(x)=\delta(x-R)$ or the potential equal
to zero except the close vicinity of space point $R$. In this
point a wavicle appears in our observed Classical World when its
bra and ket meet each other. For the quantum states $k$ with the
wave function $\psi_k$ the wavicle background contribution (see
the diagram F1.3 or the left diagrams of Fig.2) is given by:
\begin{equation}\label{21}
U(R)=|\psi_k(R)|^2F_k \quad \quad \overline U=\int_VU(R)dR=F_k
\end{equation}
These expressions have clear physical meaning. It gives the
probability to find a wavicle of a given state $k$ in a given
space point $R$. We see also that if the quantum state is occupied
$F_k\neq 0$ its bra and ket will meet certainly somewhere in the
system. For $F_k=1$ we come to the well-known Born rule to treat
$|\psi(R)|^2$ as the probability for a quantum particle to be in
the space point $R$. For many quantum states and $N$ quantum
particles that can occupy them we have obviously
\begin{equation}\label{22}
U(R)=\sum_k|\psi_k(R)|^2F_k \quad \quad \overline U=\sum_kF_k=N
\end{equation}
Thus we see that zero-phase wavicles look like classical particles
thus demonstrating corpuscular properties. However, there is a
principal difference between classical particles (i.e. material
points) and wavicles. The point-particles in our Classical World
always exist as really observable entities whereas the wavicles
appear there only after measurements (i.e. after the encounter of
their bra and ket constituents). Before such encounters the
wavicles are only potentially observable.
\par
Now let us consider the phase-dependent wavicles ($\alpha$-type
wavicles) that are responsible for peculiar quantum effects. First
of all note that we should have at least two quantum states for
their appearance (see the diagram F1.4). At least two {\it
occupied} states are needed for quantum exchange (see the diagrams
of Fig.2).
\par
The contribution of $(kp)$-wavicles (i.e. $k$-bra and $p$-ket) is
given by the non-diagonal matrix element $\langle
k|U|p\rangle=\psi_k^\dag(R)\psi_p(R)$ together with the
corresponding phase multipliers (\ref{15}) and the mean
distribution value as $\sqrt{F_kF_p}$. The $(pk)$-wavicles give
the complex-conjugated contribution.
\par
Thus for the phase-dependent $\alpha$-type wavicles the total
contribution to the observable quantity $\Delta U(R,t)$ is given
by:
\begin{equation}\label{23}
\Delta U(R,t)=\sum_{k\neq p}\psi_k^\dag(R)
\psi_p(R)e^{i\alpha_{pk}(t)}\sqrt{F_kF_p}+c.c
\end{equation}
We see here the sum of contributions of various bra+ket pairs
(wavicles with various phases). Their phases are random quantities
(see (\ref{15})). Because of symmetry between bra and ket $\Delta
U(R,t)$ is a real quantity and can be rewritten as
\begin{equation}\label{24}
\Delta U(R,t)=\sum_{k\neq
p}|\psi_k^\dag(R)\psi_p(R)|\sqrt{F_kF_p}e^{i\Phi_{kp}(R,t)}+ c.c
\end{equation}
The total phase $\Phi_{kp}(R,t)$ is the sum of $\alpha_{kp}(t)$
and the phase $\gamma_{kp}(R)$ of the product
$\psi_k^\dag(R)\psi_p(R)=|\psi_k^\dag(R)\psi_p(R)|e^{i\gamma_{kp}(R)}$.
Note that the phase $\gamma_{kp}$ is nonzero even for real space
wave functions because of their orthogonality. For two real
orthogonal functions there should be a number of space points
where their product changes its sign. In these points the phase
$\gamma$ change by $\pm\pi$. Thus $\Delta U(R,t)$ is the sum of
complex quantities $\rho e^{i\Phi}$ with $\rho>0$ and the phase
$\pm\Phi$. Note that being integrated over the system volume
$\Delta U(R,t)$ vanishes as well as it vanishes after phase
averaging. The negative parts of $\Delta U(R,t)$ require the
introduction of negative or even complex numbers in order to treat
them in a probabilistic way.
\par
Let us divide $\rho\cos\Phi$ into two parts:
\begin{equation}\label{25}
\rho\cos\Phi=\rho[\cos^2(\Phi/2)-\sin^2(\Phi/2)]\equiv
(+\rho)P+(-\rho)Q, \quad P+Q=1
\end{equation}
One part becomes the probability of positive result $+\rho$ while
other part becomes the probability of negative result $-\rho$. The
total probability remains unity as it should be. For random phase
we have mean zero probabilities:
\begin{equation}\label{26}
\overline{\cos\Phi}=\overline{\cos^2(\Phi/2)}-\overline{\sin^2(\Phi/2)}=1/2-1/2=0
\end{equation}
Now let us consider the case where phase-dependent wavicles appear
under perturbation action from initial zero-phase wavicles. As an
example take the electron of the hydrogen atom of a stationary
orbit at time t=0 (see the diagram F1.3). Then it emits or absorbs
a photon and passes to another stationary orbit. To get the final
orbit from the initial orbit two transitions are necessary
(bra-bra) and (ket-ket) (see the diagrams F1.5). Since the bra and
ket are independent objects these transitions occur randomly at
different time moments $t_1$ and $t_2$. Thus we get the
(kk)-wavicle before $t_1$ and the $(pp)$-wavicle after $t_2$. The
intermediate (pk) or (kp) wavicles with phases $\pm\Phi(t)$ exist
during random time interval $\Delta t=t_2-t_1$. They describe Bohr
"quantum jumps" between stationary orbits. These "jumps" were the
object of fierce disputes between Bohr and Schr\"{o}dinger in the
heroic time of QM. Unfortunately both used only wave-function
language (ket language) and naturally came to nothing. In the
bra+ket language the inevitability of fast "jumps" between
prolonged stationary orbits becomes evident.
\par
Now consider the action of soft potentials that unable to cause
transitions between quantum states but can change wave function
phases. Taking a constant perturbation potential $V$ which
commutes with Hamiltonian $H$ we get according to (\ref{5}):
\begin{equation}\label{27}
e^{-iHt-iVt}\psi(x,0)=e^{-iVt}\psi(x,t), \quad
e^{iHt+iVt}\psi^\dag(x,0)=e^{iVt}\psi^\dag(x,t)
\end{equation}
Neglecting switch on/off perturbations this (adiabatic) potential
during its action $\Delta t$ create additional phases $\varphi=\pm
V\Delta t$ for the bra and ket. Despite such perturbations the
wavicle phase remains the same when the perturbation acts in the
same way on its bra and ket. However, if the actions on the bra
and ket are different the wavicle will acquire the additional
phase. In the usual interference experiments a flow of zero-phase
wavicles go through two slits or two channels. Then they are
detected as screen marks or detector clicks. Thus there are two
ways for a wavicle of the flow. It can go by one or another way
thus revealing its "corpuscular nature" and creating the simple
sum of two background pictures. Alternatively it can go through
both ways (its bra goes one way while its ket another way or vice
versa). In this case the wavicle acquires a phase. Then it
participates in the constructive or distractive interference
according to its phase (see (\ref{25})). Note that interference
only redistributes background picture and does not change its
intensity. The total number of events (marks or clicks) remains
the same.
\par
The result of the two-way interference is shown in the Fig.3. The
contributions of two left diagrams describe the corpuscular
wavicle conduct when it passes one of possible two ways. Two right
diagrams show how the wavicle reveals its wave property by passing
both ways.
\begin{figure}[htb]
\begin{center}
\includegraphics[width=4in]{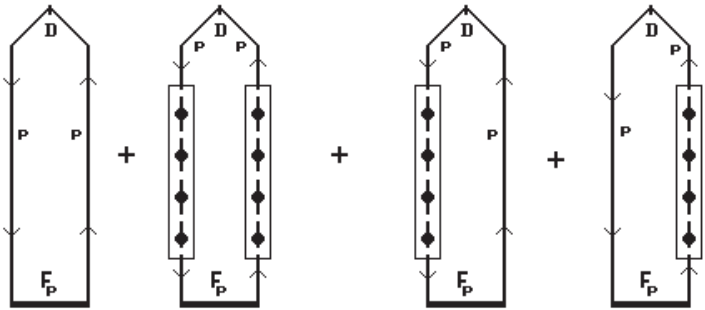}\\
\end{center} \caption{\label{phase3} Two-way interference}
\end{figure}\\
We see now that the great mystery of QM (i.e. quantum particle
self-interference) will become an obvious triviality when we
interpret the mysterious wavicle as a bra+ket pair. The necessary
but puzzling wavicle passage through two slits also becomes
natural. The use of the bra+ket language helps to make more
transparent other interference phenomena [\cite{gg19}].
\section{Phase and Second Quantization}
The second quantization is the usual and highly popular instrument
of QM mathematics. But there are no phases in the forms that we
use above. Of course, rightly used, the method can adequately
treat all quantum effects where phases play a role. However the
explicit absence of phases frequently impedes the proper
interpretation of mathematical expressions and understanding of
the physics of described phenomena. For instance, it is difficult
to see the "common cause of the past" for the quantum exchange
correlation (see the diagrams of Fig.2 and the expression
(\ref{19}). Also it is not easy to treat superconduction phenomena
without introduction of "anomalous averaging" which violates the
formal rules of second quantization.
\par
Though phases are absent in the expressions of observable
quantities they are present in the time-dependent amplitudes of
second quantization operators. The creation and annihilation
operators (better to call them the bra and ket operators) are
given by
\begin{equation}\label{28}
a^\dag(t) = e^{iHt}a^\dag e^{-iHt}=e^{i\omega t}a^\dag \quad
,\quad a(t) = e^{iHt}ae^{-iHt}=e^{-i\omega t}a
\end{equation}
where $\omega\equiv\epsilon_k$ is the energy of a given state and
the Hamiltonian has the form
\begin{equation}\label{29}
H=\sum_k\epsilon_k a^\dag_k(t) a_k(t)=\sum_k\epsilon_k a^\dag_k
a_k
\end{equation}
The initial phases in the state occupancy can be introduced
according to convention $\varphi=\omega t_0$ where $t_0$ is the
time moment when the bra and ket meet to form a wavicle.
\par
In the second quantization mathematics the lines on the diagrams
(see the figures above) correspond to the average commutators of
bra and ket operators
$\langle\Psi|[a(t_1)a^\dag(t_2)]_{\mp}|\Psi\rangle$ while wavicles
correspond to their averaged one-time correlators $\langle
\Psi|a^\dag(t)a(t)|\Psi\rangle$.
\par
Note that two-time correlators $\langle
\Psi|a^\dag(t_2)a(t_1)|\Psi\rangle$ or $\langle
\Psi|a^\dag(t_1)a(t_2)|\Psi\rangle$ represent separate bra or ket
at time $t_2>t_1$ with occupancy numbers $F$ at time $t_1$. They
have rapidly oscillating phase multipliers $e^{\pm i\omega
(t_2-t_1)}$ and do not describe observable physical quantities.
Wavicles with phases can have two different phase multipliers
$e^{i\omega t}$ and $e^{-i\omega' t}$ and oscillate with frequency
differences $\Delta\Omega=\omega-\omega'$ which can be small. In
principle they are observable and describe time-dependent
fluctuations. With implicit phase inclusion they are represented
by two-particle two-time correlators of the form
$\langle\Psi|a_1^\dag(t_1)a_1(t_2)a_2^\dag(t_2)a_2(t_1)|\Psi\rangle$.
Such correlators correspond to mean observable quantities (see,
e.g. (\ref{20})).
\section{Thermal Bath and Quantum Bath}
In the previous sections we did not distinguish the occupancy
numbers and distribution functions. In quantum mechanics one uses
the microscopic occupation numbers which for fermions are equal to
$0$ or $1$ and several units for bosons. In kinetics, however,
instead of them it is more convenient to use so-called
"coarse-grained" distribution functions which are averaged for
many adjacent quantum states or for many trials. Such functions
can have arbitrary state occupation values. Among them the most
important ones are the equilibrium distribution functions. They
are the Fermi-Dirac distribution for fermions and the
Bose-Einstein distribution for bosons as well as the Boltzman
distribution for classical particles.
\par
The equilibrium distributions do not require concrete physical
mechanisms for realizations and follow from the general
thermodynamic principle of maximal entropy. To justify the
application of this principle one should postulate the existence
of random interactions of very small intensity between the system
under consideration and its surrounding. Such interactions of
various nature constitute the thermal bath (or thermostat) which
ensures the equilibrium of the system and the relaxation to it
after various perturbations.
\par
The thermostat acts on the distribution functions which are formed
by zero-phase wavicles (i.e. the bra+ket pairs with equal
frequencies and initial phases). These wavicles represent Bohr's
stationary orbits which in many respect look as classical
particles. The wavicles formed by bra and ket with largely
different frequencies oscillate rapidly and serve as Bohr's
quantum jumps of brief duration between the stationary orbits. The
jumps occur under actions of random potentials which should be
sufficiently hard to initiate transitions between the orbits. Such
potentials are the parts of thermostat.
\par
The experience shows that bra and ket of zero-phase wavicles
prefer to keep themselves together thus looking at sufficiently
large space and time scales as single objects (e.g. photons in
light beams). The interference shows that a wavicle can change
(lose or acquire phases) under the action of soft (adiabatic)
potentials (\ref{27}) when their bra or ket suffer different
potential actions. If the potential changes equally the bra and
ket phases the wavicle phase will remain zero. Thus the
association of bra and ket of zero-phase wavicles (e.g. their
going by the same path) favors the conservation of zero-phase
wavicles while the dissociation reduces their number. Since this
number is invariant in average (\ref{22}) the processes of losing
or acquiring phases by zero-phase wavicles are similar to the
establishment of equilibrium by thermostat actions.
\par
Taking such observations into account we come to the conclusion
that phase-independent wavicles find themselves in a sort of
equilibrium as compare with phase-dependent wavicles. The always
existing random adiabatic potentials may act as a mechanism
supporting zero-phase equilibrium for wavicles. By the analogy
with thermostat (Thermal Bath) we may call these potentials by
"Quantum Bath" or "quantostat".
\section{Phase Origin}
The bra and ket space functions can be identical and real. Then
the difference between bra and ket with the same indices in this
case are the rotation multipliers $e^{\pm i\omega t}$. In the
diagrams (see Fig.1) the bra-multiplier $e^{+i\omega t}$
corresponds to the down-line arrow while the ket-multiplier
$e^{-i\omega t}$ corresponds to the up-line arrow. The arrows
means only the sign of rotation and not the evolution along or
against time direction. The evolution always is going along the
time according to the causality condition. It becomes obvious if
we include damping into the frequency (real points in diagram
lines).
\par
The time evolution of bra and ket $e^{\pm i\omega t}$ reminds the
complex solution of the harmonic oscillator equation. The
$\cos\omega t$ and $\sin\omega t$ are also solutions and they are
real. But they do not satisfy the general causality condition that
the variation of a quantity at the time $(t+0)$ is determined by
its value at the time $(t-0)$. Two exponential solutions satisfy
this condition so their use as two amplitudes in this sense is
preferable. For an oscillator the classical bra and ket amplitudes
($a^\dag$ and $a$) as well as their quantum analogues include the
coordinate and momentum parts. Therefore they describe
simultaneously the position and the velocity. The imaginary unit
permits to unite these complimentary physical quantities as a
single entity at the same time retaining their separate existence.
The phase $\phi(t)=\omega t$ reflects the ratio between these
parts in the amplitudes and implicitly also in their product:
\begin{eqnarray}\label{30}
1=e^{i\omega t}e^{-i\omega t}=(\cos\omega t + i\sin\omega
t)(\cos\omega t-i\sin\omega t)=\\\nonumber \cos^2\omega
t+\sin^2\omega t\equiv P+K=1
\end{eqnarray}
The potential energy and position correspond to the real part of
the solutions while the kinetic energy and velocity correspond to
their imaginary part. At a given moment of time two relative parts
of the total energy exist as the potential energy $P=\cos^2\omega
t$ and the kinetic energy $K=\sin^2\omega t$. One can also
consider $P$ and $K$ as the probabilities to find the energy in
its potential or kinetic forms. If we accept the indivisibility of
a single quantum then it will reveal itself in two incompatible
events with the probabilities $P+K=1$. For $N\gg 1$ quanta they
form two distinct parts as the classical potential and kinetic
energies with equal mean values.
\par
Now divide the time interval $(0T)$ by a number of parts $\Delta
t_j$. Then ignoring the state variations we can compare the
unobserved quantum evolution with the corresponding potentially
observed evolution as an equality:
\begin{eqnarray}\label{31}
1=\Pi_0^T e^{i\omega \Delta t_j}e^{-i\omega\Delta
t_j}=\Pi_0^T(\cos^2\omega \Delta t_j + \sin^2\omega\Delta
t_j)=\Pi_0^T(p_j+k_j)=1
\end{eqnarray}
At the left we see the bra and ket unobservable quantum evolution.
At the right it transforms by possible observations into a series
of incompatible events with probabilities $p_j+k_j=1$ for
complimentary quantities. One can include also a possible fast
quantum jumps between stationary orbits with different frequencies
$\omega\rightarrow\omega_j$. This way the phases implicitly govern
the results of observations.
\par
For non real space wave functions (i.e. for plain waves) one
should add the space parts of phases to the time and initial parts
of bra and ket phases.
\par
\section{Conclusion}
The use of the bra+ket language for the interpretation of QM
formulae permits to get simple and natural answers on a number of
questions which "one cannot ask" in the usual (incomplete)
ket-language. The bra and ket phases and the resulting wavicle
phase are the necessary elements of the physical picture though as
a rule they do not appear in the final expressions for the
observed physical quantities.
\par
The quanta of energy that are the basic elements of Quantum World
may be imagined as localized entities with permanent oscillations
between their complimentary (potential and kinetic) components.
The oscillations usually are too fast and because of this the
separate bra or ket are unobservable. But their combinations in
the form of wavicles have much less oscillation frequencies and
emerge in our Classical World as observed quantum particles which
actually become classical objects. The implicit phases of bra and
ket components of wavicles cause the peculiar "non-intuitive"
properties of wavicles.
\par
Let us emphasize that any adequate interpretation of QM
mathematics is impossible without the natural physical picture of
bra+ket=wavicle.
\par
One can compare the practice of getting reasonable explanations of
quantum phenomena by the incomplete ket-language as an attempt to
march using only one leg.
\par
As a result it inevitably leads to various non-physical fantasies
grossly contradicting all established principles of Physics
inherited from the past or to the capitulation "it is impossible
to understand Quantum Mechanics".

\end{document}